\documentclass[preprint,aps,prb]{revtex4}
\usepackage[dvips]{graphicx}
\usepackage{color,array,dcolumn}
\usepackage{amsmath}
\usepackage{amssymb}

\begin{document}

\title{Cohesive properties of noble metals by van der Waals-corrected Density 
Functional Theory} 

\author{Alberto Ambrosetti and Pier Luigi Silvestrelli}
\affiliation{Dipartimento di Fisica e Astronomia, 
Universit\`a di Padova, via Marzolo 8, I--35131, Padova, Italy,
and DEMOCRITOS National Simulation Center, of the Italian Istituto 
Officina dei Materiali (IOM) of the Italian National 
Research Council (CNR), Trieste, Italy}

\begin{abstract}
\date{\today}
The cohesive energy, equilibrium lattice constant, and bulk modulus 
of noble metals are computed by different 
van der Waals-corrected Density Functional Theory methods, including
vdW-DF, vdW-DF2, vdW-DF-cx, rVV10 and PBE-D. Two specifically-designed
methods are also developed in order to effectively include dynamical
screening effects: the DFT/vdW-WF2p method, based on
the generation of Maximally Localized Wannier Functions, and  
the RPAp scheme (in two variants), based on a single-oscillator model of 
the localized electron response. 
Comparison with results obtained without explicit inclusion of 
van der Waals effects, such as with the LDA, PBE, PBEsol, or the
hybrid PBE0 functional, elucidates the importance of a suitable
description of screened van der Waals interactions even in the case
of strong metal bonding. 
Many-body effects are also quantitatively evaluated within the RPAp approach.
\end{abstract}

\maketitle

\section{Introduction}
van der Waals (vdW) interactions are ubiquitous in nature and can be 
of importance even in densely packed systems, characterized by
strong ionic, covalent, or metallic bonds, where these types of 
interactions are commonly assumed to be negligible.\cite{Liu,Berland}

\begin{figure}
\centerline{
\includegraphics[width=19cm]{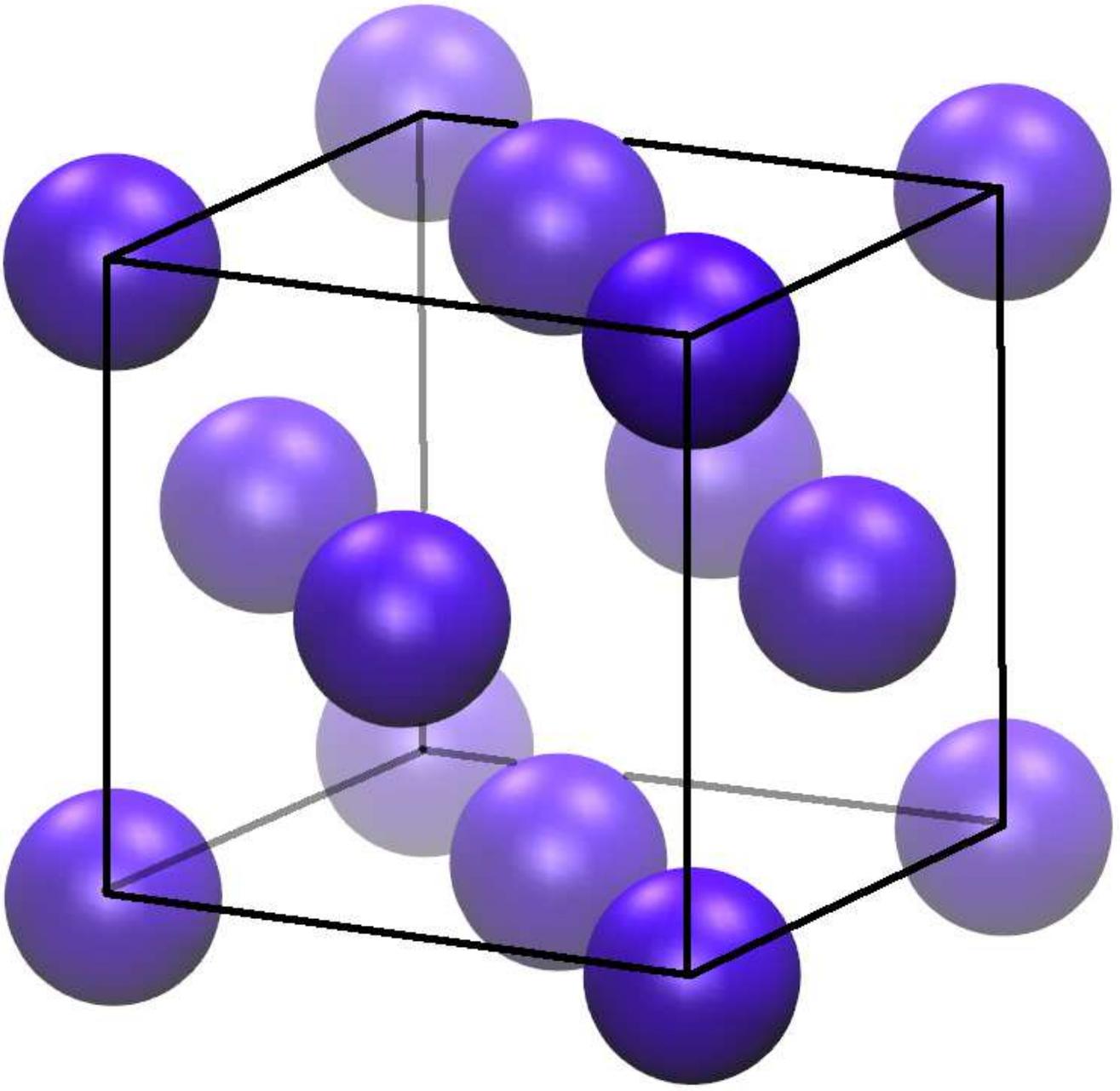}
}
\caption{Equilibrium crystal structure, face-centered-cubic (FCC), of Cu, Ag,
and Au.}
\label{figstructure}
\huge
\end{figure}

In particular, the effect of vdW forces in noble metals has been investigated
a long time ago by Rehr and Zaremba\cite{Rehr} who adopted a simplified 
model in which the ions are regarded as nonoverlapping (with the core electrons 
which are relatively well localized) and immersed in a 
uniform electron gas made by the conduction electrons.
Therefore, in spite of the hybridization between bands, the optical
properties are reasonably well described by first separating the electrons 
into polarizable core states embedded in a quasifree-electron gas.\cite{Maggs}
This corresponds\cite{Maggs} to regarding the $d$ electrons as localized 
atomic orbitals in the spirit of the tight-binding approximation, while 
treating the $s$ electrons with nearly-free-electron theory.
By taking into account the conduction-electron screening of the ion-ion 
interactions and using effective ionic polarizabilities derived from the
measured optical constants, the vdW contribution to the cohesive energy per 
atom was found\cite{Rehr} to be 0.21, 0.42, and 0.63 eV for Cu, Ag, and Au, 
respectively in the equilibrium face-centered cubic geometry (FCC,
see Fig. \ref{figstructure}).
These values should be compared to the total cohesive energies in these metals,
which are 3.50, 2.96, and 3.78 eV, respectively.
On going from Cu to Ag to Au, the increasing magnitude of the 
polarizability reflects the increasingly larger low-frequency oscillator 
strengths, with a significant part of the low-frequency oscillator strengths 
that is attributable to $d$-electron transitions.\cite{Rehr}
Interestingly, Rehr and Zaremba obtained an expression for the 
dipole-dipole contribution to the 
polarization force which is analogous to the usual expression for the 
vdW interaction between atoms in a molecular crystal.\cite{Maggs}
The presence of the electron gas gives rise to a dynamically 
screened interaction between the ions described by the dielectric function 
of the electron gas which is approximated by the random-phase approximation 
(RPA) expression evaluated for the values of the $r_s$ parameters 
corresponding to the ``free-electron'' densities of the noble metals.
In the simplest model one assumes that the core fluctuations are 
dominated by a single excited state leading to an
approximated formula for the screened vdW interactions in a polarizable
metal.\cite{Maggs} This approximation is certainly crude for a noble metal
since excitations to a continuum in the range 5-50 eV are important,
but the result demonstrates an essential qualitative point, namely that
if the dielectric constant does not show substantial momentum dependence,
dispersion forces approximately keep their usual asymptotic form even in the presence 
of dynamic screening by an electron gas.\cite{Maggs}
As expected, the screening is most effective for frequencies smaller than the 
plasma frequency.
Contributions to the polarization forces due to higher-order terms than the
dipole-dipole one are estimated to increase the vdW correction by
roughly 20\%.\cite{Rehr}
The above description applies equally to any simple metal which has a highly 
polarizable core, but whose core electrons may be well separated from the 
$s-p$ band.\cite{Maggs}

From a theoretical point of view, it is well accepted that Density Functional
Theory (DFT), which represents the most popular and widely adopted ab initio
approach for condensed-matter calculations, fail to capture vdW forces,
at least within the standard implementations based on the 
Local Density Approximation (LDA) or the Generalized Gradient 
Approximation (GGA), which are still the most used for the calculations
of structural and electronic properties of solids.
In the last years several practical methods have been proposed
to make DFT calculations able to accurately describe vdW effects at a 
reasonable computational cost (see, for instance,
refs. \onlinecite{Riley,MRS,Klimes}).
We have developed a family of such methods, all based on the generation
of the Maximally Localized Wannier Functions (MLWFs),\cite{Marzari}
successfully applied to a variety of 
systems, including small molecules, water clusters, 
graphite and graphene, water layers interacting with graphite, 
interfacial water on semiconducting substrates,
hydrogenated carbon nanotubes, 
molecular solids, and the interaction of rare gases, small molecules, 
and graphene with metal surfaces.\cite{silvprl,silvmetodo,silvsurf,CPL,silvinter,ambrosetti,Costanzo,Ar-Pb,Ambrosetti2012,C3,PRB2013,QHO-WF,QHO-WFs,Ni-gra}
Of a particular value is the possibility of dealing with metals;
in fact insulators could be somehow treated even using atom-based
semiempirical approaches where an approximately derived $R^{-6}$ term,
multiplied by a suitable short-range damping function, is explicitly 
introduced. Instead,
in our methods the atom-based point of view assumed in standard
semiempirical approaches is replaced by an electron-based point of view,
so that the schemes are also naturally applicable to systems, such as metals and
semimetals, which cannot be described in terms of assemblies of atoms only
weakly perturbed with respect to their isolated configuration.
Besides  electron-based models, effective approaches may also be
designed in such a way to separate local and non-local 
polarizability components.
Although the polarizability disentanglement is a non-trivial procedure
which requires detailed information on the electronic structure of the system, 
the introduction of a suitably parameterized atom-based 
model Hamiltonian can thus provide an alternative
pathway to the computation of vdW interactions.\cite{Bade}
We remark that in the case of vdW-corrected DFT schemes applied to metallic
systems a proper inclusion of screening effects is 
essential.\cite{Ruiz,PRB2013} 
We also stress that cross-validation between independent vdW
methods not only permits a more rigorous accuracy benchmarking, but 
also provides better insight into the relevant physical ingredients of 
the theory.

In order to unravel the role of  long-range correlation and screening
effects, we assess here the performance of different
vdW-corrected Density Functional Theory methods applied to the
description of cohesive properties of bulk noble metals.
In particular, to capture complex dynamical screening effects, 
we develop a specific version (see below) of our DFT/vdW-WF2
approach,\cite{PRB2013} and the RPAp scheme, based on a single-oscillator 
model of the localized electron response (see below). 
Moreover, we consider the dispersion corrected PBE (PBE-D\cite{PBE-D}), 
vdW-DF,\cite{Dion,Langreth07} vdW-DF2,\cite{Lee-bis} 
vdW-DF-cx,\cite{BerlandPRB,BerlandJCP} rVV10\cite{Sabatini} methods,
the TPSS\cite{TPSS} and PBE0\cite{PBE0} functionals,
and the simpler Local Density Approximation (LDA) and semilocal GGA
(in the PBE\cite{PBE} and PBESol\cite{PBEsol} flavors) approaches.
PBEsol\cite{PBEsol} is a simple revision of PBE to improve the description of
solids and surfaces.
In the PBE-D scheme DFT calculations at the PBE level are corrected
by adding empirical $C_6/R^6$ potentials with parameters derived from 
accurate quantum chemistry calculations for atoms, while in other methods,
such as vdW-DF, vdW-DF2, vdW-DF-cx, and rVV10, vdW effects are included by 
introducing DFT nonlocal correlation functionals.
In particular, vdW-DF\cite{Dion,Langreth07} denotes the first version of the
Rutgers-Chalmers functional, vdW-DF2\cite{Lee-bis} the second version which
updates both the exchange and correlation functional to improve the 
description of bonding among small molecules, 
vdW-DF-cx \cite{BerlandPRB,BerlandJCP} one of the most recent versions
based on a different exchange functional, and rVV10\cite{Sabatini} is
the revised, more efficient version of the original VV10 scheme\cite{Vydrov}
which relies on the optimization of two parameters using a reference
database.
Finally, TPSS\cite{TPSS} is a meta-GGA functional whose mathematical form 
is based on the PBE functional,\cite{Haas} while PBE0\cite{PBE0,HSE03}
denotes a hybrid functional which combines PBE with a given fraction of 
exact exchange.

\section{Screened vdW Methods}
\label{secmethods}

\begin{figure}
\centerline{
\includegraphics[width=15cm]{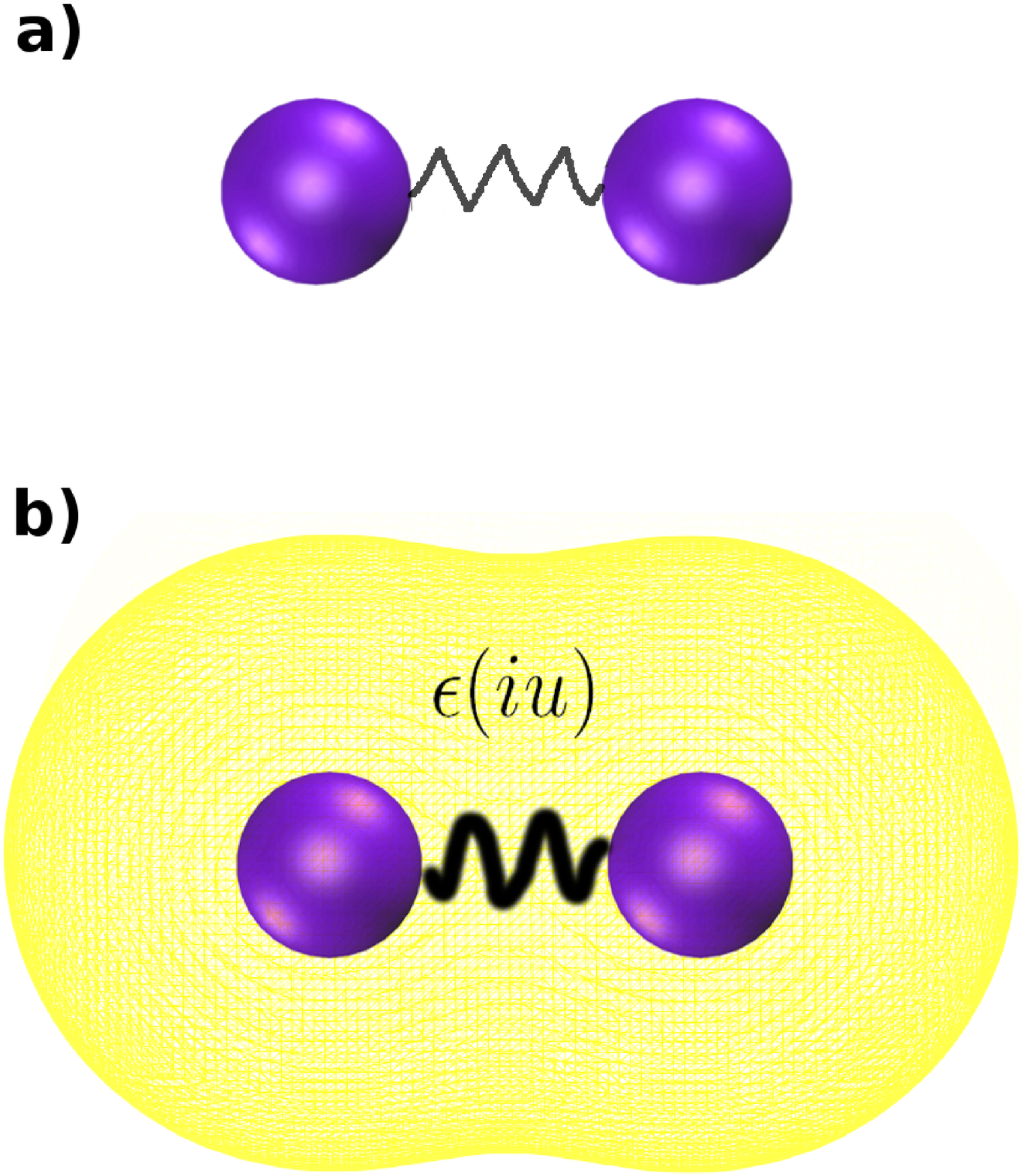}
}
\caption{Intuitive representation of screened vdW interactions in noble metals. 
In panel {\bf a)} two atoms
interact via unscreened dipole-dipole interaction, as in molecular crystals. 
In panel {\bf b)} the presence of a medium made of delocalized
electrons introduces in noble metals a dynamical screening 
of the dipole-dipole interaction, effectively described by the
dielectric function $\epsilon(iu)$.}
\label{figmethod}
\huge
\end{figure}

We propose and apply here two independent theoretical methodologies for the
computation of screened vdW energy contributions in bulk metallic systems.
Both approaches are based on a distinction between localized electrons
contributing to the local polarizability, and delocalized states, which
--in analogy to the homogeneous electron gas-- introduce a dynamical 
screening of the Coulomb interaction, as outlined in Fig. \ref{figmethod}.
The screening is effectively described through a frequency-dependent 
dielectric function, which we approximate through a single-pole expression 
(see refs. \onlinecite{Maggs,Rehr}) as
\begin{equation}
\epsilon(iu)=1+\frac{\omega_p^2}{u^2} \,,
\label{epsilon}
\end{equation}
where $\omega_p$ is the plasma frequency of the {\it electron gas}.
Then, by exploiting the adiabatic connection fluctuation-dissipation (ACFD) 
formula,\cite{jcpproof} where use is made of the 
{\it dressed} Coulomb interaction, by integration
over the frequency one obtains the dynamically screened vdW energy.

Different approaches, instead, are followed for describing the polarizability 
of localized electrons. In the first method (DFT/vdW-WF2p -- where 
``p'' indicates the introduction of plasmon pole screening) a MLWF 
partitioning of the electronic charge is introduced, which allows a 
transparent and straightforward 
separation between localized and delocalized states.
The second method (denoted as RPAp), instead is based on a 
single-oscillator description
of the localized electron response, inspired by the recent 
MBD method,\cite{jcpproof,jcpmethod} 
and permits to compute many-body dispersion effects within the RPA.

\subsection{DFT/vdW-WF2p}
Our DFT/vdW-WF2 method (see additional details in refs. 
\onlinecite{C3,PRB2013}) relies on the
well known London's expression\cite{london} where
two interacting atoms, $A$ and $B$, are approximated by 
coupled harmonic oscillators
and the vdW energy is taken to be the change of the zero-point energy
of the coupled oscillations as the atoms approach; if only a single
excitation frequency is associated to each atom, $\omega_A$, $\omega_B$,
then

\begin{equation}
E^{London}_{vdW}=-\frac{C_6^{AB}}{R^6_{AB}}=-\frac{3e^4}{2m^2}\frac{Z_A Z_B}{\omega_A \omega_B(\omega_A+\omega_B)}\frac{1}{R_{AB}^6}
\label{lond}
\end{equation}

where $Z_{A,B}$ is the total charge of A and B, and $R_{AB}$ is 
the distance between the two atoms ($e$ and $m$ are the electronic charge
and mass).

Now, adopting a simple classical theory of the atomic polarizability, 
the polarizability 
of an electronic shell of charge $eZ_i$ and mass $mZ_i$, tied to a heavy 
undeformable ion can be written as

\begin{equation}
\alpha_i\simeq \frac{Z_i e^2}{m\omega_i^2}\,.
\label{alfa}
\end{equation}

Then, given the direct relation between polarizability and atomic
volume,\cite{polvol} we assume that $\alpha_i = \gamma S_i^3$,
where $\gamma$ is a proportionality constant, so that $\alpha_i$ is
expressed in terms of the MLWF spread, $S_i$.
Recasting eq. \eqref{lond} in terms of the quantities defined above,
one obtains an explicit expression for the $C_6^{ij}$ vdW coefficient
relative to the vdW interaction between the $i$-th and the $j$-th electronic
orbitals belonging to different fragments:

\begin{equation}
C_{6}^{ij}=\frac{3}{2}\frac{\sqrt{Z_i Z_j}S_i^3 S_j^3 \gamma^{3/2}}
{(\sqrt{Z_j}S_i^{3/2}+\sqrt{Z_i}S_j^{3/2})}\,.
\label{c6}
\end{equation}

In previous applications\cite{C3,PRB2013} the proportionality constant 
$\gamma$ was set up by imposing that the exact value for
the H atom polarizability
($\alpha_H=$4.5 a.u.) is obtained. Here, a more convenient choice 
(see also below) for $\gamma$ is to impose that, by summing the 
contributions to the polarizability coming from the $d$-like MLWFs
describing the core ions, one reproduces the experimental    
effective ionic polarizabilities derived from the
measured optical constants:\cite{Rehr}
$ \alpha_{ion} = \sum_i \alpha_i = \sum_i \gamma S_i^3$.

In order to achieve a better accuracy, one must properly deal with
{\it intrafragment} MLWF charge overlap. 
This overlap affects the effective orbital volume,
the polarizability, and the excitation
frequency (see eq. \eqref{alfa}), thus leading to a quantitative effect on the
value of the $C_6$ coefficients.
We take into account the effective change in volume due to intrafragment
MLWF overlap by introducing a suitable reduction factor $\xi$
obtained by interpolating between the limiting cases of fully
overlapping and non-overlapping MLWFs (see ref. \onlinecite{C3}).
We therefore arrive at the following expression for the $C_6$ coefficient:

\begin{equation}
C_{6}^{ij}=\frac{3}{2}\frac{\sqrt{Z_i Z_j}\xi_i S_i^3 \xi_j S_j^3 \gamma^{3/2}}
{(\sqrt{Z_j\xi_i} S_i^{3/2}+\sqrt{Z_i\xi_j} S_j^{3/2})}\,,
\label{c6eff}
\end{equation}

where $\xi_{i,j}$ represents the ratio between the effective and the
free volume associated to the $i$-th and $j$-th MLWF.

Finally, the vdW interaction energy is computed as:

\begin{equation}
E_{vdW}=-\sum_{i<j}f(R_{ij})\frac{C_6^{ij}}{R^6_{ij}} \,,
\label{EvdW}
\end{equation}

where $f(R_{ij})$ is a short-range damping function which maintains
the same basic functional form adopted in previous 
applications\cite{C3,PRB2013} but it is slightly modified following the 
prescription proposed by Tkatchenko and Scheffler\cite{Tkatchenko} 
and is defined as :

\begin{equation}
f(R_{ij})=\frac{1}{1+e^{-a(R_{ij}/R_s-1)}}\,.
\end{equation}

We remark that this short-range damping function is
introduced not only to avoid the unphysical divergence of the
vdW correction at small fragment separations, but also
to eliminate double countings of correlation effects, by considering that
standard DFT approaches properly describe short-range
correlations.

The parameter $R_s$ is proportional to
the sum of the vdW radii: $R_s = s(S_i+S_j)$, with $s=0.94$ (that is a
value optimized for the underlying PBE functional\cite{Tkatchenko})
and, following Grimme {\it et al.},\cite{Grimme}
$a \simeq 20$; note that the results are only mildly
dependent on the particular value of this parameter, at least within a
reasonable range around the $a=20$ value.
Although this damping function introduces a certain degree of empiricism
in the method, we stress that $a$
is the only ad-hoc parameter present in our approach, while all the others
are only determined by the basic information given by the MLWFs, namely
from first principles calculations.
In practice, since we focus on bulk noble metals with interatomic distances
corresponding to lattice constants close to the experimental values, the
damping function defined above is effective (that is its value is
significantly smaller than 1) only for bulk Cu, since in this 
case the average sum of the spreads of 2 MLWfs belonging to adjacent Cu 
atoms (2.55 \AA) is very close to the interatomic distance (2.55 \AA, at
the experimental lattice constant), while this is not the case for
Ag and Au which are characterized by larger nearest-neighbor distances
(the literature reference data for the experimental lattice constants 
of Cu, Ag, and Au are 3.61, 4.09, and 4.08 \AA, respectively).

To get an appropriate inclusion of metal screening effects
we adopt the approximated formula\cite{Maggs} for the screened vdW interactions
in a polarizable metal (see above) based on the plasma frequency, which
consists in multiplying the $C_6^{ij}/R^6_{ij}$ contribution in 
eq. \eqref{EvdW} by the reduction factor:
$\left( w_{ij}/(w_{ij} + w_p) \right)^3$\;,
where $w_{ij}$ denotes the average excitation frequency attributed to the
$i$-th and $j$-th MLWFs, $w_{ij} = (w_i + w_j)/2$, with 
$w_i = \sqrt{Z_i/(\gamma\xi_i S_i^3)}$\;,
and $w_p$ is the plasma frequency, which can be directly related to the
metal $r_s$ parameter: $w_p \propto{r_s}^{-3/2}$. 

Therefore, in summary, our vdW-WF2 scheme specifically developed for
dealing with noble-metal bulk systems, hereafter referred to as
vdW-WF2p, differs from the previous versions
essentially in: (i) a different choice of the $\gamma$ parameter
to reproduce ionic polarizabilities, (ii) a modified damping function,
and (iii) a different implementation of the metal screening correction
(based on the plasma frequency),
that in previous applications was only aimed at describing adsorption
processes on metal substrates.

\subsection{RPAp}
Analogously to DFT/vdW-WF, RPAp is based on an effective model, where
the response of strongly bound electrons is described in terms of localized atomic polarizabilities,
interacting through a uniform medium described by the dielectric
function of eq.\eqref{epsilon}.

The localized ionic polarizabilities are mapped into the response of single quantum
Drude oscillators, coupled via dipole-dipole interaction.
The resulting effective Hamiltonian is written as\cite{jcpproof,jcpmethod}
\begin{equation}
H=-\frac{1}{2} \sum_{p=1}^{N} \nabla^2_{\boldsymbol{\mu}_p} + 
\frac{1}{2} \sum_{p=1}^{N} \omega_p^2 \boldsymbol{\mu}_p^2 + 
\sum_{p>q}^{N} \omega_p \omega_q \sqrt{\alpha^0_p \alpha^0_q} 
\boldsymbol{\mu}_p {\mathfrak{T}}_{pq} \boldsymbol{\mu}_q \,,
\label{HMBD}
\end{equation}
where each atom $p$ is characterized by a static dipole polarizability, 
$\alpha^0_p$, and a characteristic oscillator frequency, $\omega_p$, 
while ${\boldsymbol{\mu}}_p$ represents the charge displacement of atom 
$p$ from its equilibrium position, $\mathbf{R}_p$.
The dipole interaction tensor is defined as
${\mathfrak{T}}_{pq} = \nabla_{\mathbf{R}_p} \otimes \nabla_{\mathbf{R}_q} v(R_{pq})$, 
where $v(R_{pq})$ is the (frequency dependent) dynamically screened Coulomb interaction 
acting between atoms $p$ and $q$.
As in DFT/vdW-WF2p, a simple, parameter-free short range damping is introduced,
following ref. \onlinecite{jcpproof}.

While in DFT/vdW-WF2p the charge partitioning is naturally accomplished
through the use of the MLWFs, a different approach is adopted in RPAp:
following ref. \onlinecite{Rehr}, we make use of the ionic polarizability
extracted from optical data by subtracting the {\it free electron} contribution.
The ionic polarizability is then fitted through a Lorenzian expression,
corresponding to the response of a single Drude oscillator:
\begin{equation}
\alpha_{ion}(iu)=\frac{\alpha_0}{1+u^2/\omega_{0}^2}\,.
\label{qhoresp}
\end{equation}
The static polarizability $\alpha_0$ is set to the reference value of
ref. \onlinecite{Rehr}, while two alternative fitting schemes 
are proposed in order to determine the optimal oscillator frequency $\omega_0$.

As a first approach (RPAp1), $\omega_0$ is chosen in such a way to
reproduce the dipole-dipole screened dispersion energies reported in
ref. \onlinecite{Rehr} at second perturbative order.
In order to check the accuracy of the single Drude oscillator approximation,
a second approach (RPAp2) is also adopted, namely, 
$\omega_0$ is straightforwardly obtained by fitting the
frequency-dependent polarizability curves of ref. \onlinecite{Rehr} 
through eq.\eqref{qhoresp}.

Once the effective Hamiltonian is parameterized 
(see data in Table \ref{tableRPA}), the screened vdW energy
is computed at the RPA level via the ACFD formula:\cite{jcpproof}
\begin{equation}
E_{\rm RPAp}=\frac{1}{2\pi}\int_0^{\infty} {\rm du} {\rm Tr}[\ln(1-AT)]\,,
\label{acfdt}
\end{equation}
where $A$ is a diagonal matrix defined as $A_{ij}=-\alpha_{ion}(iu)\delta_{ij}$.
We stress that, at variance with refs. \onlinecite{jcpproof,jcpmethod}, 
explicit integration over frequency $u$ is necessary here, due to the 
presence of the dynamically screened Coulomb interaction.

\subsection{Computational details}
We here apply our specific version of the DFT/vdW-WF2 method described above 
(vdW-WF2p) and the RPAp scheme in the two variants (RPAp1 and RPAp2) 
to compute the basic cohesive properties, namely 
cohesive energy, equilibrium lattice constant, and bulk modulus 
of noble metals Cu, Ag, and Au.
All calculations have been performed 
with the Quantum-ESPRESSO ab initio package\cite{ESPRESSO} and the
MLWFs have been generated as a post-processing calculation using
the WanT package.\cite{WanT} 
We consider a periodically-repeated simulation cell (corresponding to the
FCC crystal structure appropriate for noble metals) containing a
single metal atom.
Electron-ion interactions were described using ultrasoft
pseudopotentials (norm-conserving for TPSS calculations) 
by explicitly including 11 valence electrons per metal atom.
As done in previous studies\cite{Ruiz2016} we adopted a $16\times16\times16$
$k$-point sampling of the Brillouin Zone.

To obtain the cohesive energy per atom, simulations were also performed
by considering isolated metal atoms contained in a large supercell, so that
to make spurious interatomic interactions negligible. To this aim 
spin-polarized calculations were carried out since
the relative spin polarization vanishes in solids around
equilibrium, but not in their free atoms,\cite{Csonka} 
given the presence of an odd number of valence electrons.
Using the DFT/vdW-WF2p approach the vdW correction to the cohesive energy 
per atom
is obtained (similarly to what done to estimate the dipole-dipole interaction
in ref. \onlinecite{Rehr}) by summing one half of the contribution 
coming from the interaction of the MLWFs reference supercell with
the periodically-repeated MLWFs (considering the periodic FCC structure), the 
sum being rapidly convergent.\cite{Ashcroft}
Calculations were carried out by varying the lattice constant from about
$-3$\% to $+3$\% with respect to the experimental equilibrium value;
then the cohesive-energy curve is fitted using the Murnaghan 
equation\cite{Murnaghan} to estimate the precise equilibrium values.   

For our DFT/vdW-WF2p calculations 
we chose the PBE\cite{PBE} reference DFT functional, which is probably
the most popular GGA functional, provides excellent results in many 
cases, and is the standard functional for solid-state calculations.
As often done in recent studies (see, for instance, ref. \onlinecite{Rosa}),
the PBE functional was also adopted to generate pseudopotentials used in
all the other vdW-corrected schemes we applied, although, in principle
one could generate pseudopotentials consistent with the specific 
functionals (for instance belonging to the vdW-DF family). Extensive
tests on many systems showed that this approach is reasonable since
the underlying approximation is small and this avoids using 
pseudopotential that are not well tested as the PBE ones.

Since we are investigating bulk interatomic distances close to
the experimental equilibrium values, in order to make the calculations
more efficient, for each system, the MLWFs were generated only 
at the experimental lattice constants and using a $8\times8\times8$
$k$-point sampling of the Brillouin Zone: in fact optimizing the MLWF
spread with several $k$ points typically leads to quite slow convergence
process; we have tested that these approximations do not introduce 
sizeable errors. 

As discussed in the Introduction section, in bulk noble metals one can
safely consider\cite{Rehr,Maggs} the $d$ electrons as localized orbitals,
while the $s$ electrons can be described with nearly-free-electron theory. 
Therefore, in our DFT/vdW-WF2p approach, the vdW-correction was
implemented by considering the contribution from the MLWFs generated 
by the optimizing process including only the narrowest 
energy window containing 5 bands (the $d$-like ones), which
represents a well-defined criterium to minimize mixing with other 
states.\cite{Souza} 

Regarding the RPAp method, due to the slow convergence of many body effects 
with respect to the system size, all calculations are performed in real space,
adopting periodic boundary conditions and making use of an extended
cubic supercell containing 864 atoms. Clearly, both the frequency 
integration and the relatively large size of the adopted supercell 
imply a non-negligible computational overload with respect to DFT/vdW-WF2p. 
The higher computational cost, however,
is justified by the possibility to estimate non-trivial energy contributions 
beyond finite-order perturbative approaches.

\section{Results and Discussion}
In Tables II-IV we report the equilibrium cohesive energy, lattice constant, 
and bulk modulus of bulk Cu, Ag, and Au, computed by using 
our DFT/vdW-WF2p and RPAp schemes, and the other different 
methods discussed above.
The {\it cohesive energy}, $E_c$, is defined as
\begin{equation}
E_c=E_{free}-E
\end{equation}
where $E_{free}$ is the total energy of the isolated free metal atom
(contained in a large supercell, see above) and $E$ is instead the 
total energy (per atom) of the bulk metal. 

\begin{figure}
\centerline{
\includegraphics[width=19cm]{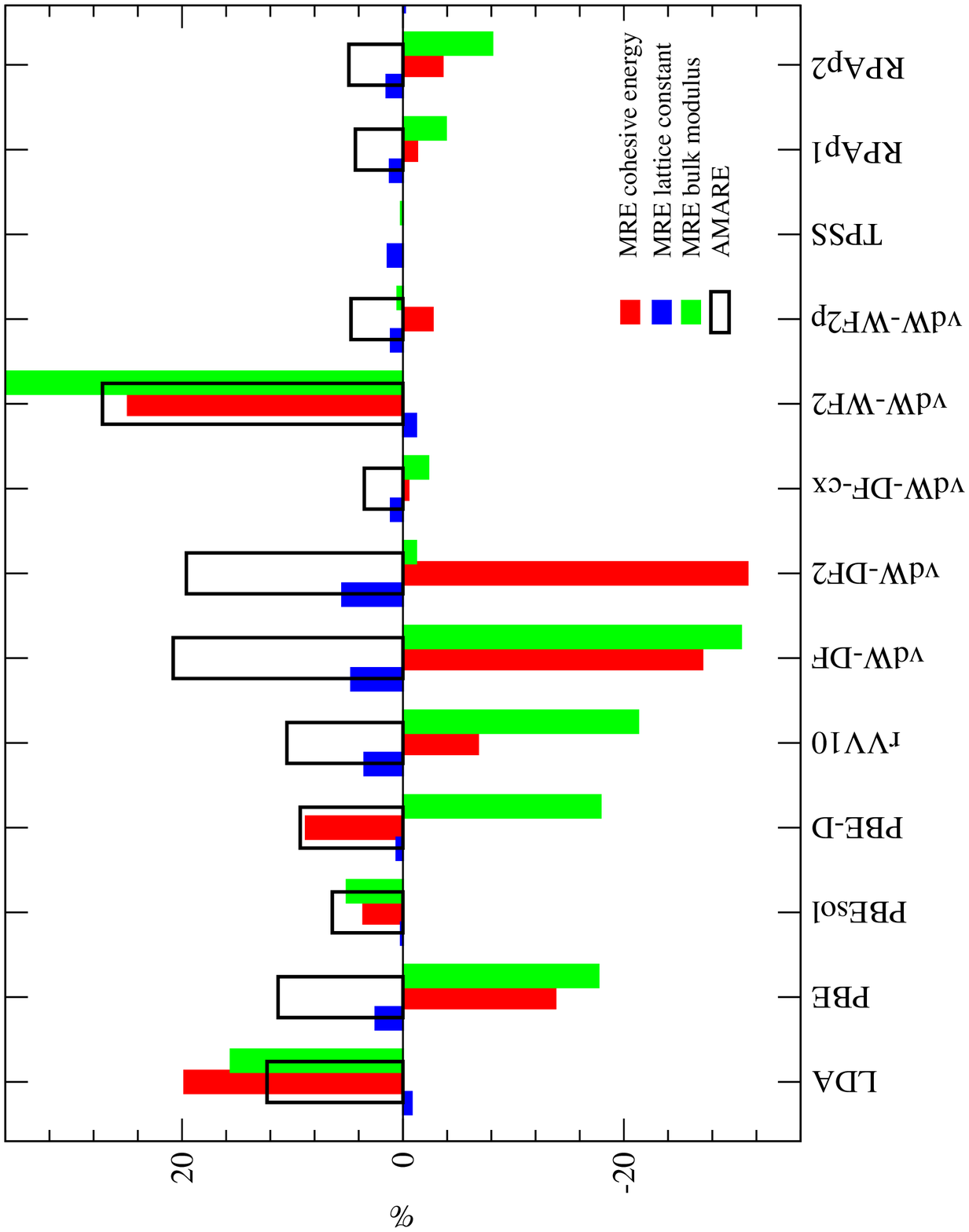}
}
\caption{Mean relative error (MRE) relative to cohesive energy,
lattice constant and bulk modulus (plain color filled bars), and
averaged mean absolute relative error (AMARE, pattern filled bars).}
\label{figMRE}
\huge
\end{figure}

In order to get a more reliable comparison with experimental results,
the zero-point phonon and finite-temperature effects should be properly 
taken into account,\cite{Haas,Csonka} since experimental quantities are 
often measured at room temperature so that they are not directly 
comparable with the results of ground-state electronic structure 
calculations performed at 0 K.
In particular, the zero-point phonon effects correspond to a zero-point 
anharmonic expansion which tends to make the optimal lattice-constant larger.
For instance, the uncorrected experimental results are closer to PBE than LDA, 
while the experimental values corrected by zero-point phonon effects are 
smaller and thus closer to LDA values.\cite{Csonka}
Moreover, temperature and phonon effects can modify the bulk
modulus up to 5\%-8\% for metals so that the corrected experimental bulk moduli
are stiffer on average.
Finally, also the experimental cohesive energies must be corrected by the 
zero-point vibration energy calculated from the Debye temperature.\cite{Csonka} 
In Table V we report the mean relative error (MRE), 
the mean absolute relative error (MARE) of the 
cohesive energy, lattice constant and bulk modulus, and the  
MARE value averaged over the 3 different quantities considered (AMARE);
the MREs and the AMARE are also plotted in
Fig. \ref{figMRE}. As can be seen,
the equilibrium lattice constant is reasonably reproduced by all the methods,
while the same is not true for the cohesive energy and the bulk modulus
which are much more sensitive to the chosen approach.

First considering the LDA and (GGA) PBE functionals, one can see that 
both give errors 
in the estimated cohesive energy and bulk modulus of comparable magnitude
(AMARE= 12.3\% and 11.3\%, respectively)
though of opposite sign (see Fig. \ref{figMRE}), as already noticed in the 
literature (see, for instance, refs. \onlinecite{Haas,Csonka}): 
LDA systematically 
overestimates cohesive energy and bulk modulus while PBE underestimates
these quantities. Moreover LDA slightly underestimates the lattice constant
while PBE (more significantly) overestimates it. 
Clearly, neither LDA nor PBE turns out to be adequate for an accurate
description of bulk noble metals. 
The (GGA) PBEsol functional,\cite{PBEsol} characterized by a modification 
of the exchange contribution of the original PBE to better describe solids, 
evidently improves considerably the performances with respect to 
PBE (AMARE= 6.4\%), although a tendency to overestimate particularly 
the cohesive energy and the bulk modulus can be observed, 
again in line with previous findings.\cite{Csonka} 
Notice that LDA, and the (GGA) PBE and PBEsol functionals by construction
cannot properly describe vdW effects.  
Since the PBE functional reduces to LDA for homogeneous electron densities,
in principle the metallic electrons should be already accurately described 
within PBE. However, while this is probably true for simple metals
like Na (PBE underestimates the cohesive energy by less than 5\%, see
ref. \onlinecite{Csonka}), the application to  noble metals, 
characterized by a large fraction (10/11) of relatively localized $d$-like 
electrons, shows that PBE misses a significant portion of the interaction
leading to an underestimate of the cohesive energy and bulk modulus and an
overestimate of the lattice constant.
We can classify this neglected part of the interaction as ``vdW interaction'',
although, vdW-corrected PBE calculations often tend to overbind because of
an overestimate of the long-range part of the
exchange contribution.\cite{silvsurf,ambrosetti,Dion,Ruiz2016} 

The PBE-D functional, including vdW correction by a semiempirical 
approach,\cite{PBE-D} is only marginally superior (AMARE= 9.3\%) to LDA
and PBE, but worse than PBEsol; in fact, it overestimates the cohesive 
energy (as LDA) and underestimates the bulk modulus (as PBE).

In principle, seamless vdW-corrected DFT methods, based on genuine
nonlocal functionals, should perform better than PBE-D, particularly
for metals, where an empirical, atom-based vdW correction is expected
to be not adequate. However, as can be seen from Tables II-IV and Fig. \ref{figMRE},
different schemes, belonging to this class of approaches, perform quite 
differently for bulk noble metals. 
In fact, rVV10 (AMARE= 10.5\%) is worse than PBE-D and does not even 
significantly improve over PBE, by clearly overestimating the lattice constants
and underestimating the cohesive energy and (above all) the bulk modulus.
The situation is even worse for vdW-DF and vdW-DF2 (AMARE= 20.8\% and 19.6\%,
respectively), where the errors involved in the estimate of the
equilibrium lattice constants (too large) and cohesive energies (too small) 
are particularly evident.
Instead, the vdW-DF-cx variant\cite{BerlandPRB,BerlandJCP} of the vdW-DF
family clearly represents a remarkable improvement for describing the
basic properties of the noble metals, with AMARE= 3.5\%, the agreement 
with experimental data being particularly striking for Ag.

In order to assess the relevance of a proper description of dynamical screening 
effects in noble metals we now turn to the specifically developed approaches 
presented in Section \ref{secmethods}.
Remarkably, our DFT/vdW-WF2p, RPAp1, and RPAp2 methods exhibit performances 
similar to vdW-DF-cx (AMARE= 4.7\%, 4.3\%, and 4.9\%, respectively), 
again giving excellent results for Ag and improving over PBEsol.  
By comparison of the DFT/vdW-WF2p data with the same
quantities evaluated without taking
the conduction-electron screening into account (for brevity this method is
denoted as DFT/vdW-WF2, see Tables II-IV and Fig. \ref{figMRE}), it is clear
that the cohesive energy is substantially overestimated by the latter 
approach, which also predicts a too large bulk modulus and underestimates
the equilibrium lattice constant (AMARE= 27.2\%).
This is clearly due to the fact that the vdW-correction of eq. \eqref{EvdW}
is too large if the reduction factor $\left( w_{ij}/(w_{ij} + w_p) \right)^3$
is not included, as confirmed by Table VI, where the vdW contribution to the
cohesive energy, $E_{vdW}$, defined as the difference between the 
equilibrium cohesive energy of bulk Cu, Ag, and Au, and the same quantity 
computed with the PBE functional, is listed; 
the results are compared with the estimates reported by Rehr 
and Zaremba.\cite{Rehr} Note that the ``dipole-dipole 
contribution to the polarization force'' of ref. \onlinecite{Rehr} should be
directly comparable with, for instance, the $E_{vdW}$ quantity computed by
our DFT/vdW-WF2p method, since it is analogous to the 
usual expression for the vdW interaction between atoms in a molecular 
crystals:\cite{Maggs} particularly for Ag and Au the agreement is rather good.
In line with ref. \onlinecite{Rehr}, if the conduction-electron 
screening is neglected, the vdW contributions are 2-3 times larger than 
the screened ones; the same is true for the semiempirical PBE-D scheme
which clearly overestimates the cohesive energy, particularly for Au
(see Tables II and VI). 
Note that reduction factors similar to ours, are suggested
in ref. \onlinecite{Ruiz} to correct effective C$_6$ coefficients
in noble metals, by including screening effects estimated by the
Lifshitz-Zaremba-Kohn theory.\cite{LZK}   

Interestingly, the RPAp method allows for a direct estimate of many-body 
contributions\cite{jcpl} to the cohesive energy in bulk noble metals.
In fact, by performing a Taylor series expansion of eq.\eqref{acfdt},
truncated to second order, one obtains the following expression:
\begin{equation}
E^{(2)}_{\rm RPAp}=-\frac{1}{4\pi}\int_0^{\infty} {\rm du} {\rm Tr}[(AT)^2]\,,
\label{acfdt2}
\end{equation}
which mathematically corresponds\cite{jcpproof} to the pairwise
summation of two-body energy contributions. The quantity
$E^{\rm MB}_{\rm RPAp} = E_{\rm RPAp} - E^{(2)}_{\rm RPAp}$ hence includes all
many-body effects, {\it i.e.} all the energy contributions beyond the 
two-body component.
By evaluating $E^{\rm MB}_{\rm RPAp}$
we find that many-body effects induce a significant 
reduction of the vdW cohesive energy and smaller changes in the other
quantities: for instance, in the case of Ag, fitting
the cohesive-energy curve obtained using \eqref{acfdt2} gives
for the vdW contribution 0.446 eV/atom, for the equilibrium lattice 
constant 4.087 \AA, and for the bulk modulus 1079 kbar, so that 
(compare with RPAp1 data in Tables III, IV, and VI)
many-body effects lead to variations in these quantities 
by -5.2\%, +0.2\%, and -1.2\%, respectively (once again the equilibrium
lattice constant appears to be the least affected quantity). 
We also observe, comparing RPAp1 to RPAp2, that the first methodology,
where the Hamiltonian of Eq. \eqref{HMBD} 
is parameterized in such a way to reproduce the effective {\it screened}
interatomic $C_6$ coefficients of ref. \onlinecite{Maggs}, 
is slightly more accurate than the second,
where the $C_6$ coefficients turn out to be slightly underestimated.
In any case this further supports the validity of the present approach,
since different parameterizations lead to limited differences on 
the overall performances of the methods, that are in both
cases similar to those obtained by DFT/vdW-WF2p.

Coming back to the good performances of the PBEsol DFT functional reported
above, which are only slightly worse than those of the vDW-corrected
DFT/vdW-WF2p, RPAp, and vdW-DF-cx methods, 
the improvement of PBEsol with respect
to PBE can be explained by the fact that PBEsol restores\cite{PBEsol} 
the density-gradient expansion appropriate for slowly varying densities 
(as those characterizing the valence-electron densities in densely packed 
solids), while PBE instead is optimized to reproduce the exchange energy
of free atoms. Therefore, in PBEsol the revised exchange contribution 
turns out to be almost equivalent to the nonlocal vdW contribution which
is added to PBE in the DFT/vdW-WF2p approach (or included in the nonlocal
vdW-DF-cx functional).

As can be seen from Tables III, IV, and V, the meta-GGA TPSS 
functional\cite{TPSS} performs quite well (in agreement with ref. 
\onlinecite{Csonka}) for estimating the equilibrium lattice constant and 
bulk modulus, however a full assessment of the method cannot be given,
since, due to numerical problems in the spin-polarized 
calculations for the isolated atoms, the cohesive-energy estimate 
is not available (the same is true also in ref. \onlinecite{Csonka}).
  
Finally, data obtained\cite{HSE03} at the PBE0 level, namely using 
an hybrid functional which contains a certain amount of exact exchange
and is much more computationally expensive than the other considered methods,
are listed in Tables II,III, and IV for Cu and Ag.
The evident underestimate of the cohesive energies (14\% and 22\% for Cu, and
Ag, respectively) and of the bulk modulus values together with the significant
overestimate of the equilibrium lattice constant, clearly show that 
hybrid functionals such as PBE0 (or the almost equivalent HSE03 which 
allows to evaluate the short-range Fock operator on a coarser
$k$ mesh than PBE0\cite{HSE03}) do not include vdW effects properly, as already 
pointed out elsewhere.\cite{Ruiz2016}

\section{Conclusions}
In conclusion, we have computed the cohesive energy, equilibrium lattice constant, 
and bulk modulus of noble metals by different van der Waals-corrected Density Functional Theory methods,
including vdW-DF, vdW-DF2, vdW-DF-cx, rVV10 and PBE-D.
In addition, an extension of the DFT/vdW-WF2 (DFT/vdW-WF2p) scheme, 
based on the generation of Maximally Localized Wannier Functions,
and an effective atom-based RPA (RPAp) approach (in two variants)
have been specifically designed to treat the complex dynamical screening induced
by delocalized $s$ electrons.
Comparison with results obtained without explicit inclusion of
van der Waals effects, such as with the LDA, PBE, PBEsol, or the
hybrid PBE0 methods, elucidates the importance of a suitable
description of screened van der Waals effects interactions even in the case
of strong metal bonding. This is particularly true for cohesive energy and
bulk modulus, while lattice constants are typically already well reproduced
at the LDA/GGA level.
We confirm that the GGA PBEsol functional significantly improves on PBE.
On the whole, the vdW-DF-cx, RPAp1, and DFT/vdW-WF2p methods give
the best results.
In particular, the good performances of the DFT/vdW-WF2p and RPAp schemes 
give further support to the simple plasma-frequency model, 
proposed by Maggs and Ashcroft\cite{Maggs} for describing the  screened 
vdW interactions in a polarizable metal. In fact, while conventional 
pairwise vdW methods show a clear tendency to overestimate cohesion,
 the dynamical screening acts by effectively reducing the vdW
interaction, consistently improving the overall accuracy. 
As estimated via the RPAp method, many body effects beyond the two-body
contributions induce a further reduction of the overall vdW cohesion.
Due to the interplay between different energetic and screening contributions, 
the cohesion of noble metals hence emerges as a challenging problem,
whose accurate modeling demands the introduction of appropriate
vdW-corrected DFT approaches.


\vfill
\eject

\begin{table}
\caption{Fitted static polarizabilities and oscillator frequencies 
(in a.u.) adopted in the RPAp1 and RPAp2 methods (see text).}
\begin{center}
\begin{tabular}{|l|c|c|c|}
\hline
method           &   Cu  &   Ag  &   Au   \\ \tableline
\hline
$\alpha_0$       & 11.7  & 14.6  & 19.5   \\  
$\omega_0$ RPAp1 & 0.58  & 0.84  & 0.80  \\
$\omega_0$ RPAp2 & 0.46  & 0.75  & 0.68  \\
\hline
\end{tabular}
\end{center}
\label{tableRPA}
\end{table}
\vfill
\eject

\begin{table}
\caption{Equilibrium cohesive energy $E_c$ (in eV/atom) of bulk Cu, Ag, 
and Au, using different methods.
Results are compared with available experimental reference data
corrected by finite-temperature and zero-point effects (see text).}
\begin{center}
\begin{tabular}{|l|c|c|c|}
\hline
method          &   Cu  &   Ag  &   Au   \\ \tableline
\hline
LDA             & 4.353 & 3.612 & 4.357  \\  
PBE             & 3.323 & 2.513 & 3.042  \\
PBEsol          & 3.895 & 3.088 & 3.682  \\
\hline
PBE-D           & 3.721 & 3.078 & 4.456  \\
rVV10           & 3.392 & 2.835 & 3.533  \\
vdW-DF          & 2.732 & 2.189 & 2.694  \\
vdW-DF2         & 2.574 & 2.038 & 2.515  \\
vdW-DF-cx       & 3.636 & 3.075 & 3.754  \\
\hline
PBE0$^a$        & 3.046 & 2.329 &  ---  \\
\hline
DFT/vdW-WF2     & 4.589 & 3.469 & 4.874  \\
DFT/vdW-WF2p    & 3.697 & 2.808 & 3.523  \\
\hline
RPAp1           & 3.636 & 2.937 & 3.586  \\
RPAp2           & 3.569 & 2.879 & 3.466  \\
\hline
 expt.          & 3.524 & 2.972 & 3.810  \\
\hline
\end{tabular}
\tablenotetext[1]{ref.\onlinecite{HSE03}.}
\end{center}
\label{table2}
\end{table}
\vfill
\eject

\begin{table}
\caption{Equilibrium lattice constant $a_0$ (in \AA) of bulk Cu, Ag, and Au,
using different methods.
Results are compared with available experimental reference data
corrected by finite-temperature and zero-point effects (see text).}
\begin{center}
\begin{tabular}{|l|c|c|c|}
\hline
method          &   Cu  &   Ag  &   Au   \\ \tableline
\hline
LDA             & 3.553 & 4.020 & 4.053  \\  
PBE             & 3.676 & 4.163 & 4.173  \\
PBEsol          & 3.585 & 4.063 & 4.099  \\
\hline
PBE-D           & 3.615 & 4.155 & 4.026  \\
rVV10           & 3.699 & 4.219 & 4.211  \\
vdW-DF          & 3.754 & 4.262 & 4.249  \\
vdW-DF2         & 3.788 & 4.276 & 4.286  \\
vdW-DF-cx       & 3.627 & 4.103 & 4.108  \\
\hline
TPSS            & 3.607 & 4.136 & 4.127  \\
\hline
PBE0$^a$        & 3.636 & 4.142 &  ---  \\
\hline
DFT/vdW-WF2     & 3.599 & 3.975 & 3.996  \\
DFT/vdW-WF2p    & 3.641 & 4.088 & 4.108  \\
\hline
RPAp1           & 3.646 & 4.094 & 4.113  \\
RPAp2           & 3.658 & 4.106 & 4.128  \\
\hline
 expt.          & 3.595 & 4.056 & 4.062  \\
\hline
\end{tabular}
\tablenotetext[1]{ref.\onlinecite{HSE03}.}
\end{center}
\label{table3}
\end{table}
\vfill
\eject

\begin{table}
\caption{Equilibrium bulk modulus $B$ (in kbar) of bulk Cu, Ag, and Au,
using different methods.
Results are compared with available experimental reference data
corrected by finite-temperature and zero-point effects (see text).}
\begin{center}
\begin{tabular}{|l|c|c|c|}
\hline
method          &   Cu  &   Ag  &   Au   \\ \tableline
\hline
LDA             & 1715  & 1386  & 1945   \\  
PBE             & 1286  &  912  & 1430   \\
PBEsol          & 1662  & 1196  & 1739   \\
\hline
PBE-D           & 1332  &  650  & 1825   \\
rVV10           & 1229  &  867  & 1377   \\
vdW-DF          & 1042  &  799  & 1206   \\
vdW-DF2         & 1080  & 1434  & 1746   \\
vdW-DF-cx       & 1495  & 1105  & 1701   \\
\hline
TPSS            & 1688  & 1040  & 1698   \\
\hline
PBE0$^a$        & 1300  &  868  &  ---   \\
\hline
DFT/vdW-WF2     & 2646  & 1473  & 2843   \\
DFT/vdW-WF2p    & 1581  & 1071  & 1727   \\
\hline
RPAp1           & 1511  & 1067  & 1649   \\
RPAp2           & 1433  & 1035  & 1570   \\
\hline
 expt.          & 1420  & 1090  & 1965   \\
\hline
\end{tabular}
\tablenotetext[1]{ref.\onlinecite{HSE03}.}
\end{center}
\label{table4}
\end{table}
\vfill
\eject

\begin{table}
\caption{Mean relative error (MRE) and, in parenthesis, 
mean absolute relative error (MARE) relative to cohesive energy,
lattice constant and bulk modulus, and, in square parenthesis, 
MARE value averaged on the 3 different quantities considered (AMARE),
using different methods, listed in order of decreasing performance 
(increasing AMARE value).}
\begin{center}
\begin{tabular}{|l|c|c|c|c|}
\hline
method          & $E_c$ & $a_0$ & $B$  & AMARE \\ \tableline
\hline
vdW-DF-cx       & -0.5 (2.7) &  1.1 (1.1) & -2.3 (6.7) &  [3.5]  \\
RPAp1           & -1.3 (3.4) &  1.2 (1.2) & -3.9 (8.2) &  [4.3]  \\
DFT/vdW-WF2p    & -2.7 (6.0) &  1.1 (1.1) &  0.5 (7.1) &  [4.7]  \\
RPAp2           & -3.6 (4.5) &  1.5 (1.5) & -8.1 (8.7) &  [4.9]  \\
PBEsol          &  3.6 (5.9) &  0.2 (0.5) &  5.1(12.8) &  [6.4]  \\
PBE-D           &  8.8 (8.8) &  0.6 (1.2) &-17.9(17.9) &  [9.3]  \\
rVV10           & -6.8 (6.8) &  3.5 (3.5) &-21.3(21.3) & [10.5]  \\
PBE             &-13.8(13.8) &  2.5 (2.5) &-17.7(17.7) & [11.3]  \\
LDA             & 19.8(19.8) & -0.8 (0.8) & 15.6(16.3) & [12.3]  \\  
vdW-DF2         &-31.2(31.2) &  5.5 (5.5) & -1.2(22.2) & [19.6]  \\
vdW-DF          &-27.1(27.1) &  4.7 (4.7) &-30.6(30.6) & [20.8]  \\
DFT/vdW-WF2     & 24.9(24.9) & -1.2 (1.3) & 55.4(55.4) & [27.2]  \\
TPSS            &   ---      &  1.4 (1.4) &  0.2(12.3) &  ---    \\
\hline
\end{tabular}
\end{center}
\label{table5}
\end{table}
\vfill
\eject

\begin{table}
\caption{vdW contribution, $E_{vdW}$, in eV/atom, 
(in percentage with respect to the total cohesive energy in parenthesis),
defined as the difference 
between the equilibrium cohesive energy of bulk Cu, Ag, and Au, and
the same quantity computed with the PBE functional. 
Results are compared with the estimates by Rehr and Zaremba\cite{Rehr}
(in square parenthesis the dipole-dipole contribution to the
polarization force which is analogous to the usual expression for the
vdW interaction between atoms in a molecular crystal\cite{Maggs}).}
\begin{center}
\begin{tabular}{|l|c|c|c|}
\hline
method                   &   Cu  &   Ag  &   Au   \\ \tableline
\hline
PBE-D                    & 0.398 (11\%) & 0.565 (18\%) & 1.414 (32\%)  \\
\hline
DFT/vdW-WF2              & 1.123 (25\%) & 0.809 (24\%) & 1.731 (36\%)  \\
DFT/vdW-WF2p             & 0.374 (10\%) & 0.295 (11\%) & 0.481 (14\%)  \\
\hline
RPAp1                    & 0.313  (9\%) & 0.424 (14\%) & 0.544 (15\%)  \\
RPAp2                    & 0.246  (7\%) & 0.366 (13\%) & 0.424 (12\%)  \\
\hline
 ref. \onlinecite{Rehr}  & 0.21   (6\%) & 0.42  (14\%) & 0.63  (17\%)  \\
 ref. \onlinecite{Rehr}  &[0.179  (5\%)]&[0.347 (12\%)]&[0.524 (14\%)] \\
\hline
\end{tabular}
\end{center}
\label{table6}
\end{table}
\vfill
\eject

\begin{thebibliography}{9}
\bibitem{Liu} W. Liu, J. Carrasco, B. Santra, A. Michaelides, M. Scheffler,
              A. Tkatchenko,
              Phys. Rev. B {\bf 86}, 245405 (2012).
\bibitem{Berland} K. Berland, V. R. Cooper, K. Lee, E. Schr\"oder, 
                  T. Thonhauser, P. Hyldgaard, B. I. Lundqvist,
                  Rep. Prog. Phys. {\bf 78}, 066501 (2015).
\bibitem{Rehr} J. J. Rehr, E. Zaremba, W. Kohn,
               Phys. Rev. B {\bf 12}, 2062 (1975).
\bibitem{Maggs} A. C. Maggs, N.W. Ashcroft,
                Phys. Rev. Lett. {\bf 59}, 113 (1987).
\bibitem{Riley}  K. E. Riley, M. Pito\v{n}\'{a}k, P. Jure\v{c}ka, 
                 P. Hobza,  Chem. Rev. {\bf 110}, 5023 (2010). 
\bibitem{MRS} A. Tkatchenko, L. Romaner, O. T. Hofmann, E. Zojer, 
              C. Ambrosch-Draxl, and M. Scheffler, 
              MRS Bulletin, {\bf 35}, 435 (2010).
\bibitem{Klimes} J. Klime\v{s}, A. Michaelides, 
                 J. Chem. Phys. {\bf 137}, 120901 (2012).
\bibitem{Marzari} N. Marzari and D. Vanderbilt,
                  Phys. Rev. B {\bf 56}, 12847 (1997).
\bibitem{silvprl}  P. L. Silvestrelli,  
                   Phys. Rev. Lett {\bf 100}, 053002 (2008).
\bibitem{silvmetodo} P. L. Silvestrelli,  
                     J. Phys. Chem. A {\bf 113}, 5224 (2009).
\bibitem{silvsurf} P. L. Silvestrelli, K. Benyahia, S. Grubisi\^{c}, 
                   F.  Ancilotto, F. Toigo,  
                   J. Chem. Phys. {\bf 130}, 074702 (2009).
\bibitem{CPL} P. L. Silvestrelli,
              Chem. Phys. Lett. {\bf 475}, 285 (2009). 
\bibitem{silvinter} P. L. Silvestrelli, F.  Toigo, F.  Ancilotto, 
                    J. Phys. Chem. C {\bf 113}, 17124 (2009).
\bibitem{ambrosetti} A. Ambrosetti, P. L. Silvestrelli, 
                     J. Phys. Chem. C {\bf 115}, 3695 (2011).
\bibitem{Costanzo} F. Costanzo, P. L. Silvestrelli, Francesco Ancilotto,
                   J. Chem. Theory Comp. {\bf 8}, 1288 (2012);
                   Archives of Metallurgy and Materials {\bf 57}, 1075 (2012). 
\bibitem{Ar-Pb} P. L. Silvestrelli, A. Ambrosetti, S. Grubisi\^{c}, 
                and F. Ancilotto,
                Phys. Rev. B {\bf 85}, 165405 (2012).
\bibitem{Ambrosetti2012} A. Ambrosetti, F. Ancilotto, P. L. Silvestrelli,
                         J. Phys. Chem. C {\bf 117}, 321 (2013).
\bibitem{C3} A. Ambrosetti, P. L. Silvestrelli,
             Phys. Rev. B {\bf 85}, 073101 (2012).
\bibitem{PRB2013} P. L. Silvestrelli and A. Ambrosetti,
                  Phys. Rev. B {\bf 87}, 075401 (2013).
\bibitem{QHO-WF} P. L. Silvestrelli,
                 J. Chem. Phys. {\bf 139}, 054106 (2013).
\bibitem{QHO-WFs} P. L. Silvestrelli, A. Ambrosetti,
                  J. Chem. Phys. {\bf 140}, 124107 (2014).
\bibitem{Ni-gra} P. L. Silvestrelli, A. Ambrosetti,
                 Phys. Rev. B {\bf 91}, 195405 (2015).
\bibitem{Bade} W. L. Bade,
               J. Chem. Phys. {\bf 27}, 1280 (1957).
\bibitem{Ruiz} V. G. Ruiz, W. Liu, E. Zojer, M. Scheffler, A. Tkatchenko,
               Phys. Rev. Lett. {\bf 108}, 146103 (2012);
               W. Liu, A. Tkatchenko, and M. Scheffler, 
               Acc. Chem. Res. {\bf 47}, 3369 (2014).
\bibitem{PBE-D} S. Grimme, 
                J. Comp. Chem. {\bf 27}, 1787 (2006);
                V. Barone, M. Casarin, D. Forrer, M. Pavone, M. Sambi, 
                A. Vittadini,
                J. Comp. Chem. {\bf 30}, 934 (2009).
\bibitem{Dion}  M. Dion, H. Rydberg, E. Schr\"oder, D. C. Langreth,
                B. I. Lundqvist,  Phys. Rev. Lett.
                {\bf 92}, 246401 (2004);
                G. Roman-Perez, J. M. Soler,
                Phys. Rev. Lett. {\bf 103}, 096102 (2009).
\bibitem{Langreth07} T. Thonhauser, V. R. Cooper, S. Li,  
                     A. Puzder, P. Hyldgaard, D. C. Langreth, 
                     Phys. Rev. B {\bf 76}, 125112 (2007).
\bibitem{Lee-bis} K. Lee, \'E. D. Murray, L. Kong, B. I. Lundqvist, and 
                  D. C. Langreth, 
                  Phys. Rev. B {\bf 82}, 081101(R) (2010).
\bibitem{BerlandPRB} K. Berland, P. Hyldgaard,
                     Phys. Rev. B {\bf 89}, 035412 (2014).
\bibitem{BerlandJCP} K. Berland, Calvin A. Arter, Valentino R. Cooper, 
                     Kyuho Lee, Bengt I. Lundqvist, Elsebeth Schr\"oder, 
                     T. Thonhauser, and Per Hyldgaard,
                     J. Chem. Phys. {\bf 140}, 18A539 (2014).
\bibitem{Sabatini} R. Sabatini, T. Gorni, S. de Gironcoli,
                   Phys. Rev. B {\bf 87}, 041108(R) (2013).
\bibitem{TPSS} J. Tao, J. P. Perdew, V. N. Staroverov, and G. E. Scuseria, 
               Phys. Rev. Lett. {\bf 91}, 146401 (2003).
\bibitem{PBE0} C. Adamo, V. Barone,
               J. Chem. Phys. {\bf 110}, 6158 (1999).
\bibitem{PBE} J. P. Perdew, K. Burke, M. Ernzerhof,
              Phys. Rev. Lett. {\bf 77}, 3865 (1996).
\bibitem{PBEsol} John P. Perdew, Adrienn Ruzsinszky, Gabor I. Csonka, 
                 Oleg A. Vydrov, Gustavo E. Scuseria, Lucian A. Constantin, 
                 Xiaolan Zhou, and Kieron Burke,
                 Phys. Rev. Lett. {\bf 100}, 136406 (2008).
\bibitem{Vydrov} O. A. Vydrov, T. Van Voorhis,
                 J. Chem. Phys. {\bf 133}, 244103 (2010).
\bibitem{Haas} P. Haas, F. Tran, P. Blaha,
               Phys. Rev. B {\bf 79}, 085104 (2009).
\bibitem{HSE03} J. Heyd, G. E. Scuseria, M. Ernzerhof,
                J. Chem. Phys. {\bf 118}, 8207 (2003);
                J. Paier, M. Marsman, K. Hummer, G. Kresse, I. C. Gerber,
                J. G. Angyan,
                J. Chem. Phys. {\bf 124}, 154709 (2006). 
\bibitem{jcpproof} A. Tkatchenko, A. Ambrosetti, R.A. DiStasio Jr.,
                   J. Chem. Phys. {\bf 138}, 074106 (2013).
\bibitem{jcpmethod} A. Ambrosetti, A.M. Reilly, R.A. DiStasio Jr., A. Tkatchenko
                    J. Chem. Phys. {\bf 140}, 18A508 (2014).
\bibitem{london} R. Eisenhitz, F. London,
                 Z. Phys. {\bf 60}, 491 (1930).
\bibitem{polvol} T. Brink, J. S. Murray, P. Politzer,
                 J. Chem. Phys. {\bf 98}, 4305 (1993).
\bibitem{Tkatchenko} A. Tkatchenko, M. Scheffler,
                     Phys. Rev. Lett. {\bf 102}, 073005 (2009). 
\bibitem{Grimme} S. Grimme, J. Antony, T. Schwabe, C. M\"uck-Lichtenfeld, 
                 Org. Biomol. Chem.  {\bf 5}, 741 (2007);
                 S. Grimme, J. Antony, S. Ehrlich, H. Krieg,
                 J. Chem. Phys. {\bf 132}, 154104 (2010).
\bibitem{ESPRESSO} P. Giannozzi, S. Baroni, N. Bonini, M. Calandra, R. Car, 
                   C. Cavazzoni, D. Ceresoli, G. L. Chiarotti, M. Cococcioni, 
                   I. Dabo, A. Dal Corso, S. Fabris, G. Fratesi, 
                   S. de Gironcoli, R. Gebauer, U. Gerstmann, C. Gougoussis, 
                   A. Kokalj, M. Lazzeri, L. Martin-Samos, N. Marzari, 
                   F. Mauri, R. Mazzarello, S. Paolini, A. Pasquarello,
                   L. Paulatto, C. Sbraccia, S. Scandolo, G. Sclauzero, 
                   A. P. Seitsonen, A. Smogunov, P. Umari, R. M. Wentzcovitch, 
                   J.Phys.: Condens. Matter {\bf 21}, 395502 (2009),
                   http://arxiv.org/abs/0906.2569. 
\bibitem{WanT} WanT code by A. Ferretti, B. Bonferroni, A. Calzolari, and 
               M. Buongiorno Nardelli, http://www.wannier-transport.org ;
               see also A. Calzolari, N. Marzari, I. Souza and 
               M. Buongiorno Nardelli, Phys. Rev. B {\bf 69}, 035108 (2004).
\bibitem{Ruiz2016} W. Liu, V. G. Ruiz, G.-X. Zhang, B. Santra, X. Ren,
                   M. Scheffler, A. Tkatchenko,
                   New J. Phys. {\bf 15}, 053046 (2013);
                   V. G. Ruiz, W. Liu, A. Tkatchenko,
                   Phys. Rev. B {\bf 93}, 035118 (2016).
\bibitem{Csonka} Gabor I. Csonka, John P. Perdew, Adrienn Ruzsinszky,
                 Pier H. T. Philipsen, Sebbastien Lebegue, Joachim Paier, 
                 Oleg A. Vydrov, and Janos G. Angyan, 
                 Phys. Rev. B {\bf 79}, 155107 (2009).
\bibitem{Ashcroft} See, for instance: N. W. Ashcroft and N. D. Mermin,
                   Solid State Physics, Holt-Saunders International Editions,
                   Philadelphia, 1976.
\bibitem{Murnaghan} F. D. Murnaghan,
                    Proc. Natl. Acad. Sci. U.S.A.  {\bf 30}, 244 (1944).
\bibitem{Rosa} Marta Rosa, Stefano Corni, and Rosa Di Felice,
               Phys. Rev. B {\bf 90}, 125448 (2014).
\bibitem{Souza} I. Souza, N. Marzari, D. Vanderbilt,
                Phys. Rev. B {\bf 65}, 035109 (2001).
\bibitem{LZK} E. M. Lifshitz,
              Soviet Physics JETP {\bf 2}, 73 (1956);
              E. Zaremba and W. Kohn,
              Phys. Rev. B {\bf 13}, 2270 (1976).
\bibitem{jcpl} A. Ambrosetti, D. Alf\`{e} , R.A. DiStasio Jr., A. Tkatchenko,
               J. Phys. Chem. Lett. {\bf 5}, 849 (2014).
\end{thebibliography}
\end{document}